\documentclass[%
mathleft,%
]{an}
\usepackage{graphicx}
\usepackage{times}
\begin{document}

% The following seven commands are intended for editorial usage and
% should be ignored by the author(s).
\Pagespan{1}{}% Document's page range. 
% If second parameter is left empty, the last page is computed
% automatically.
\Yearpublication{2012}%
\Yearsubmission{2012}%
\Month{9}%   
\Volume{999}%  
\Issue{99}% 
% \DOI{This.is/not.aDOI}% 

\title{The origin of the chemical elements in cluster cores}

\author{J. de Plaa\inst{1}\fnmsep\thanks{Corresponding author:
  \email{j.de.plaa@sron.nl}}
% Example for footnote, note the usage of the \texttt{fnmsep} command
% as separator between institute number and footnote mark}
}
\titlerunning{The origin of the chemical elements in cluster cores}
\authorrunning{J. de Plaa}
\institute{
SRON Netherlands Institute for Space Research, 
Sorbonnelaan 2, 3584 CA Utrecht
}

\received{2 July 2012}
\accepted{XXXX}
\publonline{XXXX}


\abstract{%
  Metals play a fundamental role in ICM cooling processes in cluster 
  cores through the emission of spectral lines. But when and how were 
  these metals formed and distributed through the ICM? The X-ray band 
  has the unique property of containing emission lines from all elements 
  from carbon to zinc within the 0.1$-$10 keV band. Using XMM-Newton, 
  the abundances of about 11 elements are studied, which contain valuable 
  information about their origin. Most elements were formed in type Ia 
  and core-collapse supernovae, which have very different chemical yields. 
  Massive stars and AGB stars also contribute by providing most of the 
  carbon and nitrogen in the ICM. Because feedback processes suppress 
  star formation in the cluster centre, the element abundances allow us 
  to directly probe the star formation history of the majority of stars 
  that are thought to have formed between $z=2-3$. The spatial distribution 
  in the core and the evolution with redshift also provide information 
  about how these elements are transported from the member galaxies to 
  the ICM. I review the current progress in chemical enrichment studies 
  of the ICM and give an outlook to the future opportunities provided 
  by XMM-Newton's successors, like Astro-H. 
  }

\maketitle

\section{Introduction}

After the `Big Bang', the baryonic component of the Universe mainly consisted
of hydrogen and helium with traces of lithium and beryllium. The first metals 
with a higher atomic weight were produced in the first generation of stars, also
referred to as Population III stars, which started the epoch of re-ionization. 
The nature of this stellar population is very uncertain, but it likely consisted 
of intermediate-mass and high-mass stars (Vangioni et al. 2011)\nocite{vangioni2011}. 
Based on WMAP measurements, these stars formed around $z\sim 10$, when the age of the Universe
was about 500 Myr. Although these stars were the first to enrich the surrounding
gas, the total contribution to the current day chemical composition is thought 
to be small, only 10$^{-4}$ Z$_{\mathrm{solar}}$ (Matteucci \& Calura 2005)\nocite{matteucci2005}.  

The bulk of the enrichment probably occurred around $z\sim 2-3$ through the supernova
explosions following major star bursts. A compilation
of the universal star formation rate measurements as a function of redshift 
(Hopkins \& Beacom 2006) \nocite{hopkins2006} shows that the star formation rate 
peaked around $z\sim 2-3$ and declined slowly to low redshift. However, these
are averaged rates over both cluster and field galaxies. At a similar redshift, 
the Intra-Cluster Medium (ICM) in clusters of galaxies starts to form, which, 
together with feedback from Active Galactic Nuclei (AGN), quenched the star 
formation in these objects. Therefore, in the cluster of galaxies case, the star 
formation rate drops much faster than average (Gabor et al. 2010)\nocite{gabor2010}.
In that respect, clusters of galaxies are a special environment, because their 
enrichment is dominated by the products from the main star bursts at $z\sim 2-3$. 

Ferrara et al. (2005) \nocite{ferrara2005} have estimated that a major fraction 
($>$90\%) of the produced elements around $z=3$ contributes to the enrichment of 
gas in a hot phase, instead of ending in newly formed stars. The ICM in clusters 
therefore contains the bulk of the metals produced in the cluster member galaxies.
In the period of the major star bursts, the main enrichment mechanisms that were 
transporting the metals from the galaxies to the surrounding medium were galactic
winds driven by the supernova explosions and AGN uplifting of galactic gas.
At a lower level, also other enrichment mechanisms play a role. Metals can be 
ejected from the galaxies by galaxy-galaxy interactions, by sloshing motions 
of the hot ICM, and ram-pressure stripping of in-falling galaxies (see Schindler
\& Diaferio 2008 for a review\nocite{schindler2008}).

\section{Origins of elements}

The bulk of the elements heavier than beryllium are produced in the end phases
of stellar evolution. The elements in the mass range between oxygen and silicon
are mainly produced by core-collapse supernovae, because the elements with masses
higher than silicon that are produced in the core region of the massive star are 
compressed into the neutron star or black hole that is formed during the supernova 
event. Only the outer layers containing the lighter elements are ejected into the
surrounding medium. The elements from silicon to nickel, however, are the main
products of type Ia supernovae, which are exploding white dwarfs in binary systems. 
During such a supernova event, elements up to nickel are produced by explosive 
fusion and the white dwarf is disintegrated, which releases all the products in 
the surrounding medium.
 
A few light elements, however, have a different origin. Carbon is thought to be mainly
produced by massive stars during their lifetime and ejected through the stellar
winds. Nitrogen is ejected by intermediate-mass stars in the Asymptotic Giant Branch (AGB) 
phase. Since elements heavier than nickel are not produced by fusion, and are therefore
less abundant, we do not discuss the origin of these metals in this paper.

\begin{figure}[t]
\includegraphics[width=0.7\columnwidth,angle=-90]{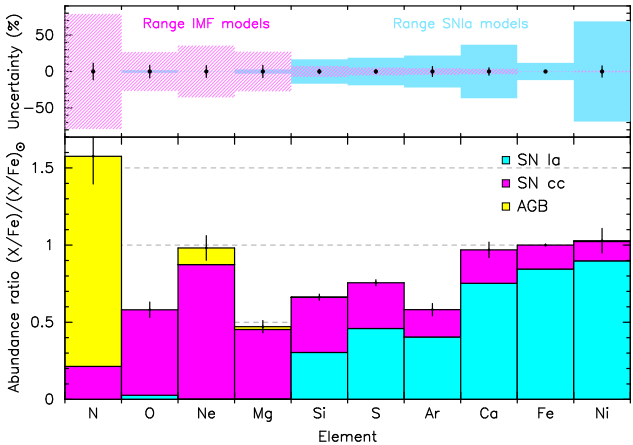}
\caption{Expected abundances measured in a 120 ks XMM-Newton observation of 
S\'ersic 159-03 (bottom panel), which is a typical bright local cluster. 
The statistical error bars were obtained from de Plaa et al. (2006)\nocite{deplaa2006}.
The estimates for the SNIa, SNcc, and AGB contributions are based on a sample of 22 
clusters (de Plaa et al. 2007)\nocite{deplaa2007} and two elliptical galaxies (Grange
et al. 2011)\nocite{grange2011}. The top panels show the typical range in SNIa and IMF
models with respect to the statistical error bars in the observation.}
\label{fig:barplot}
\end{figure}

Figure~\ref{fig:barplot} shows the expected contributions of the most abundant metals that
can be detected using the XMM-New\-ton observatory. The bars indicate the relative
contribution of each supernova type to the abundance of the elements. The 
fractions depend on the supernova models used. The effect of uncertainties in the supernova 
models and in the Initial-Mass Function (IMF) are indicated in the top panel.

\subsection{Type Ia supernovae}

Although type Ia supernovae are used as standard candles for cosmology, their progenitor
and explosion mechanism are still poorly known. The common misconception is that a
type Ia supernova occurs when a white dwarf goes over the Chandrasekhar mass.
Instead, the explosive carbon fusion in the star is ignited just before it reaches this limit. 
In recent years, optical searches for supernovae have yielded hundreds of observations of 
type Ia's which revealed a surprising variety in their properties. It has proved to be very 
difficult to link the observed supernovae to possible progenitors (see e.g. Howell 2011\nocite{howell2011} 
for a recent review).

Currently, roughly three main progenitors are being considered. The first is a `classical'
type Ia or also called `Single Degenerate' (SD), where the companion of the white dwarf is a 
main sequence or red-giant star that accretes material on the white dwarf. The second is 
a scenario where two white dwarfs merge and the less massive white dwarf is accreted onto the
more massive one, which is known as the `Double Degenerate' (DD) scenario. And the last is a 
sub-Chandrasekhar channel, where a thick layer of helium builds up on the white dwarf's 
surface either by hydrogen burning or by a helium-rich donor. Because of the lower density 
of the last systems, they are thought to produce more intermediate-mass elements like silicon, 
sulfur, and calcium than during the deflagration in the typical SD and DD cases.
   
Observations have shown that relatively luminous type Ia supernovae tend to occur in spiral 
galaxies, while the sub luminous are mainly found in elliptical galaxies with old stellar 
populations (e.g. Howell 2001\nocite{howell2001}; Sullivan et al. 2006\nocite{sullivan2006}). 
This appears to indicate that the brighter 'prompt' supernovae explode early
($<$400 Myr) after the star burst, while the sub-luminous occur with a delay of a few Gyr.
This can be explained in the DD scenario by the fact that it takes a longer time to evolve a 
star into a low-mass white dwarf compared to a high-mass white dwarf. The delay time
explanation is also supported by high-redshift studies that show that at $z=1$ type Ia supernovae
are 12\% more luminous and have less intermediate mass elements in their spectra than local 
SNIa's (e.g. Howell et al. 2007\nocite{howell2007}; Sullivan et al. 2009\nocite{sullivan2009}). 
However, also differences in initial metallicities or other unknown parameters could 
play a role. 

\begin{figure}
\includegraphics[width=\columnwidth]{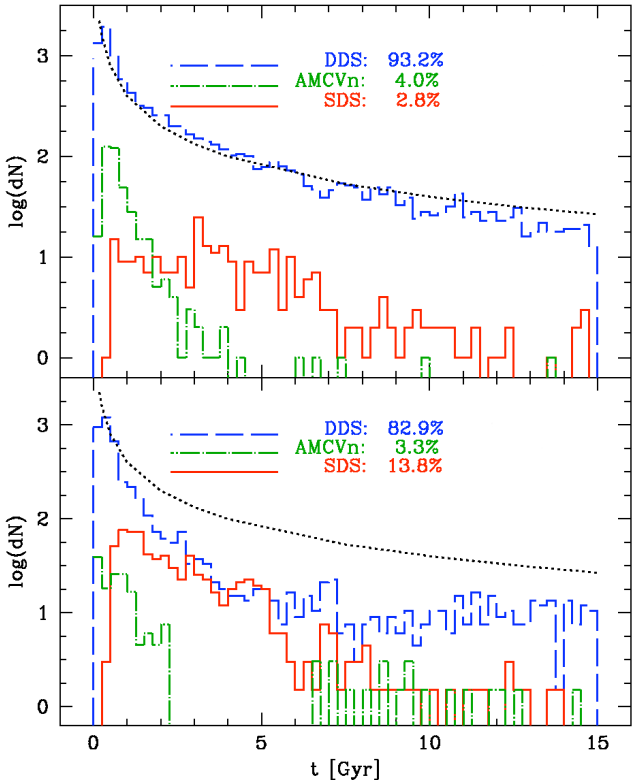}
\caption{Supernova type Ia delay time distributions for three different progenitor channels:
white-dwarf mergers (DDS), Single-Degenerate Systems (SDS), and AM CVn systems as described in 
Ruiter et al. (2009). AM CVn systems are ultra-compact systems with a WD accretor and a He-rich 
donor. The top and bottom panel represent the result from two choices of initial 
parameters, but both are calculated for stellar populations in elliptical galaxies. The dotted 
line shows the observed decline of the type Ia supernova rate of $\sim t^{-1}$, where $t$ is the 
time passed since the initial star burst. The decline is 
consistent with the DDS progenitor model.}
\label{fig:ruiter}
\end{figure}

One of the few successes to combine theory and observations in this field is the explanation
of the observed Delay-Time Distribution (DTD) using stellar population synthesis studies. 
Supernova rate measurements in elliptical galaxies (and clusters of galaxies) show that the
rate declines with $\sim t^{-1}$ after the initial star burst. Model calculations show the 
same level of decline for DD mergers, which is an indication that this channel is dominating
the SNIa rate at larger delay times (see Fig~\ref{fig:ruiter}; Ruiter et al. 2009\nocite{ruiter2009}; Mennekens et al. 
2010\nocite{mennekens2010}).
    
Due to the uncertainty in the nature of the SNIa progenitor and the poorly understood physics 
of the explosion (during the explosion, the burning front probably switches from being subsonic
to supersonic), there is a large range of supernova Ia models predicting the metal yields. Well
known examples are Iwamoto et al. (1999)\nocite{iwamoto1999} and models by Bravo et al. 
(2012)\nocite{bravo2012}. The predicted yields for various elements can vary substantially
for every model, which means that accurate abundance measurements can help to constrain the
SNIa models.

\subsection{Core-collapse supernovae}
 
Although there is also a range of different models for core-collapse supernova yields in the 
literature (e.g. Woosley \& Weaver, 1995\nocite{woosley1995}; Tsujimoto et al., 1995\nocite{tsujimoto1995};
Chieffi \& Limongi, 2004\nocite{chieffi2004}), the main parameters that determine the abundances
measured are initial metallicity and the choice of the IMF. In order to determine the total 
contribution of core-collapse supernovae to the cluster enrichment, the model yields for each
mass bin need to be weighted with the IMF in the following way 
(Tsujimoto et al, 1995\nocite{tsujimoto1995}):
\begin{equation}
M_{i} = \frac{\int_{10 ~ \mathrm{M}_{\sun}}^{50 ~ \mathrm{M}_{\sun}} ~ M_i(m) ~ m^{-(1+x)} ~ \mathrm{d}m}{
        \int_{10 ~ \mathrm{M}_{\sun}}^{50 ~ \mathrm{M}_{\sun}} ~ m^{-(1+x)} ~ \mathrm{d}m},
\label{eq:imf}	
\end{equation} 
where $M_i(m)$ is the $i$th element mass produced in a main-sequence star of mass $m$.
A standard Salpeter IMF corresponds to $x=1.35$, but this equation can, of course, be altered 
to represent other IMFs.

\subsection{Intermediate-mass AGB stars}

Since intermediate-mass stars in their AGB phase are a major source of nitrogen, it is 
important to include them in the model as well. Since the yields depend on the initial
mass of the main sequence star, we need to weigh the yields also with the IMF like in
Equation~\ref{eq:imf}. AGB star yields for different masses and initial metallicities
are calculated by, for example, Karakas (2010)\nocite{karakas2010}. Together with the 
information from the core-collapse models, the AGB stars provide an additional constraint
to the initial metallicity and IMF of the stellar population.

\section{Abundance measurements in X-rays}

\begin{figure}[t]
\includegraphics[width=\columnwidth]{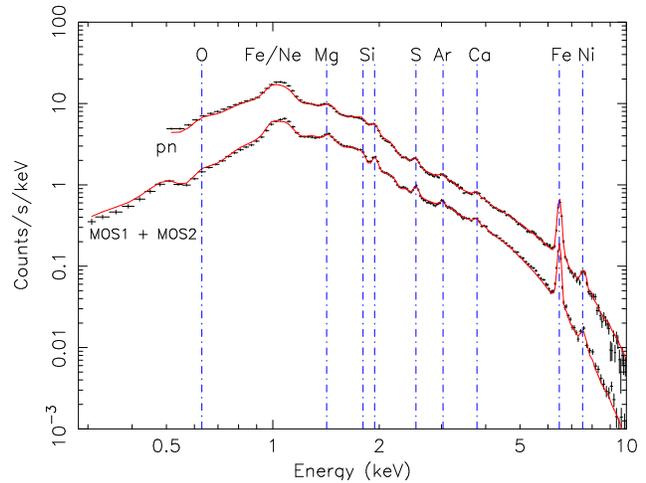}
\caption{EPIC spectrum of the cluster 2A 0335+096 with an exposure time of 130 ks. 
Adapted from Werner et al. (2006b)\nocite{werner2006}.}
\label{fig:werner}
\end{figure}

The soft X-ray band between 0.1 and 10 keV is very suitable to measure abundances in hot 
plasmas, because it contains spectral lines of all elements from carbon to nickel. The 
hot ICM in clusters is particularly interesting, because the plasma is in collisional-ionization
equilibrium, which simplifies the determination of the abundances. Moreover, the cluster
ICM contains the integrated yield of billions of supernovae, providing a general picture of 
supernova yields, contrary to studies of a few individual supernovae through their remnants in 
our galaxy. The sensitivity of XMM-New\-ton, and also Suzaku, allows us to measure abundances
with enough accuracy to constrain supernova models. An example of a deep EPIC observation of 
the cluster 2A 0335+096 in Figure~\ref{fig:werner} shows the position and strength of the
lines of the most abundant elements. The availability of these excellent data sets has triggered
a number of successful abundance studies (see also the review by Werner et al. 2008\nocite{werner2008}).    
  
\subsection{Enrichment in local clusters}

The first attempt to link measured abundances in the hot ICM in local clusters ($z<0.2$) 
was performed using the ASCA satellite (Mushotzky et al., 1996)\nocite{mushotzky1996}. Its 
instruments allowed the accurate detection of O, Ne, Mg, Si, S, Ar, Ca, Fe, and Ni for the 
first time. From this and later ASCA studies (e.g. Fukazawa et al. 1998\nocite{fukazawa1998};
Finoguenov et al. 2000\nocite{finoguenov2000b}) a general picture emerged of an ICM which
was enriched early in the cluster evolution by core-collapse supernova and later by delayed
type Ia supernovae.

\begin{figure}
\includegraphics[width=\columnwidth]{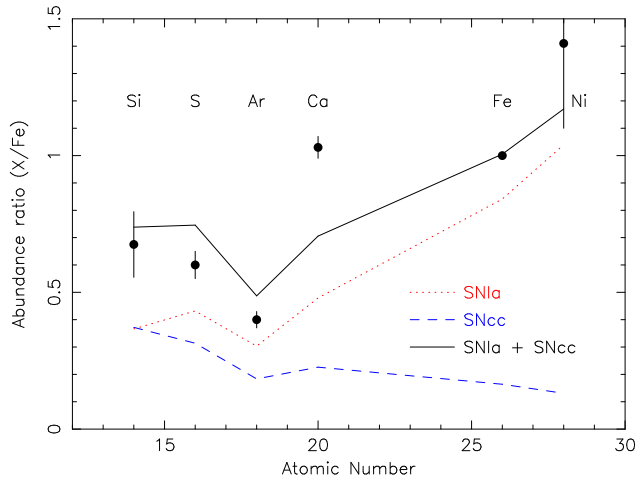}
\caption{Abundance ratios fitted with supernova yields from the WDD2 SNIa model (Iwamoto et al. 1999)
and a SNcc model with an initial metallicity $Z=0.02$ and a Salpeter IMF. The calcium abundance appears
to be underestimated. It can not explain the Ar/Ca ratio measured in this XMM-Newton sample
of 22 clusters. (Adapted from de Plaa et al. 2007).}
\label{fig:calcium}
\end{figure}

When more deep observations of clusters with XMM-Newton became available, de Plaa et al. 
(2007)\nocite{deplaa2007} performed an abundance study in a sample of 22 clusters with EPIC. 
With the obtained accuracy, it was possible to estimate the relative contributions of 
type Ia and core-collapse supernova to the enrichment of the cluster by linearly combining
the supernova yields for each type. For the most commonly used type Ia models (Iwamoto et al. 
1999\nocite{iwamoto1999}), this results in a SNIa contribution of about 30\%, which is 
remarkably similar to the SNIa/SNcc ratio of $\sim 0.2-0.4$ as observed in the optical band 
(e.g. Horiuchi et al. 2011\nocite{horiuchi2011}) given the fact that de Plaa et al. (2007) 
ignored galactic evolution effects. Surprisingly, the calcium abundance did not fit
the Iwamoto et al. (1999) models well (see Fig~\ref{fig:calcium}), while a model by 
Bravo et al. (1996)\nocite{bravo1996}, which also fitted to the Tycho supernova remnant 
(Badenes et al. 2006\nocite{badenes2006}), resulted in a much better fit of the Ar/Ca ratio 
(de Plaa et al. 2007). This showed that X-ray spectra enable us to constrain type Ia 
supernova models.

Several groups have performed similar studies with XMM-New\-ton and Suzaku data since then, but
usually on fewer clusters and with varying numbers of elements. In a small sample, Sato et al.
(2007)\nocite{sato2007} performed the same fit as de Plaa et al. (2007) using abundances measured 
with Suzaku XIS. Unfortunately, argon and calcium were not included in the fit, but their main
conclusion was consistent with de Plaa et al. (2007) considering the uncertainty introduced by 
the limited number of measured elements. Recently, Bulbul et al. (2012)\nocite{bulbul2012} used 
a modified APEC model that is able to fit the supernova type Ia to core-collapse ratio directly 
to the spectra. They also report a type Ia contribution of 30--40\%, which is consistent with 
previous work and optical data. It should be noted that this ratio depends highly on the assumed 
supernova models (De Grandi \& Molendi, 2009\nocite{degrandi2009}), but it is reassuring that
different groups find similar numbers independently. 

\subsection{The abundance of nitrogen and carbon}

In recent years, deep XMM-Newton observations of giant elliptical galaxies also allowed the 
detection of carbon and nitrogen. These two elements are not produced in large quantities in
core-collapse supernovae, but in a variety of sources. The main origins of these elements are 
still subject of debate (see e.g. Romano et al. 2010\nocite{romano2010}), but likely candidates
are metal-poor massive stars and intermediate-mass stars in their AGB phase. 

Since carbon and nitrogen have a relatively low atomic mass, their K-shell transitions are 
located in the soft X-ray band below 0.5 keV. It also causes the lines to be stronger at lower
temperatures $< 1$ keV (at temperatures of a few keV, a larger fraction of the carbon and nitrogen is 
fully ionised and not producing line emission). To detect the lines, it is best to observe 
cool elliptical galaxies with the Reflection Grating Spectrometer (RGS) aboard XMM-Newton. 
This instrument has sufficient effective area at low energies and a high spectral resolution to 
resolve the lines of these elements. Unfortunately, noise in the RGS CCD in the band where the 
carbon line is located prevents an accurate carbon measurement for weak sources.  

Werner et al. (2006a)\nocite{werner2006b} and Grange et al. (2011\nocite{grange2011}) 
analysed RGS data of the giant elliptical galaxies M87 (Werner), NGC 5044, and NGC 5813 (Grange)
and detected nitrogen. They found that the nitrogen abundances in these objects are indeed much 
higher than expected based on supernova models. From the high N/O ratio, it can be deduced that
nitrogen should still be produced in low- and intermediate-mass stars contrary to supernovae, 
because the star-formation rate in elliptical galaxies has declined very rapidly after the star 
bursts around $z\sim 2-3$.  

\subsection{Intermediate redshift clusters}

\begin{figure}
\includegraphics[width=\columnwidth]{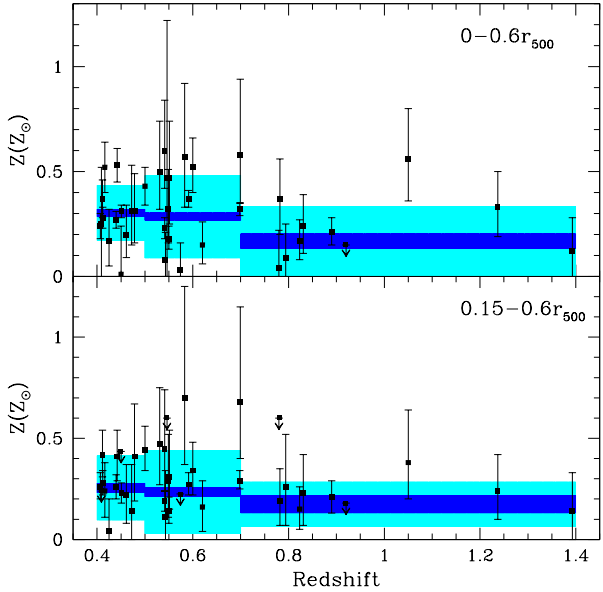}
\caption{Measured abundances versus redshift for a sample of intermediate redshift clusters.
The top panel shows the abundances measured up to 0.6R$_{500}$ and the bottom panel the results
when the core of the cluster is ignored. Adapted from Baldi et al. (2012)\nocite{baldi2012}.}
\label{fig:baldi}
\end{figure}

Studying the chemical enrichment of clusters above $z=0.2$ becomes increasingly difficult,
because of the lower flux of the sources. The Fe-K feature is usually the only feature strong 
enough to resolve at higher redshifts. Therefore, studies at these large distances focus on the
iron abundance evolution as a function of redshift. In a sample of 56 clusters with redshifts in
the 0.3$<z<$1.3 range observed with XMM-Newton and Chandra, Balestra et al. (2007)\nocite{balestra2007}
found a significant decreasing trend of the iron abundance with redshift from roughly 0.4 solar 
at $z=0.3$ to 0.2 solar at $z=1.3$. It was, however, also clear that the scatter on the measured 
data points was relatively large. In a different sample of 39 clusters observed with XMM-Newton,
Baldi et al. (2012)\nocite{baldi2012} recently reported that they could not confirm the trend
found by Balestra et al. (2007), because of the high level of scatter in the data points (see
Figure~\ref{fig:baldi}). Removing the cool cores in the spectral analysis just had a limited 
effect on the amount of scatter and did not allow them to draw firm conclusions. Since these
measurements still only sample the tail of the SNIa delay-time distribution that originates at
$z=2-3$, a strong trend may not be expected. Clearly, instruments with a much higher effective 
area are needed to reduce the statistical errors on these measurements.

\section{Discussion and future prospects}

X-ray spectroscopy of clusters of galaxies has proved to be a useful tool to study the 
enrichment history of the hot ICM, but the field is still in its early stages of development.
Using current instruments, it is possible to measure the abundance of about 10 elements
with an accuracy of 20--30\% if one includes systematic errors (De Grandi \& Molendi, 2009).
This appears to be a substantial systematic error, however, the uncertainty in the supernova
yields, given by the spread in supernova models, is sometimes more than a factor of two for 
certain elements. This means that even with these systematic uncertainties, the measurements 
can put constraints on supernova models and the IMF.

There are two main sources of systematic errors in abundance measurements: the atomic data
in spectroscopic codes and instrument calibration issues. A third one can be introduced when 
the multi-temperature structure in the gas is not taken properly into account in the spectral 
fit (Buote \& Fabian 1998\nocite{buote1998}). Both the systematics in the atomic data and the 
calibration depend very much on the specific transition and wavelength of the line. The silicon 
abundance, for example, is mainly determined by measuring the Si XIV lines at about 2.0 keV, 
which is very close to the Au-edge of the mirror and Si-edge of the detector. This is a very 
challenging wavelength band to calibrate accurately. In this case, the Si XIV lines are well 
characterized in the atomic database, while there are also other lines from, for example, 
Fe XVII that have a much larger uncertainty in their oscillator strengths (de Plaa et al., 
2012\nocite{deplaa2012}). It is very important to understand the systematics for each measured 
element individually to assess the quality of the supernova fits. 

In general, most of the measured abundances have a systematic uncertainty that is lower than or 
not much higher than 20\%, which is still much better than the scatter in the models. To beat 
down these systematic errors, investments have to be made in the development of spectral codes 
and their underlying atomic data. In addition, a high level of calibration accuracy is to be 
pursued to improve the abundance accuracy both for current and future missions.
 
\begin{figure}
\begin{center}
\includegraphics[width=0.36\textwidth,angle=-90]{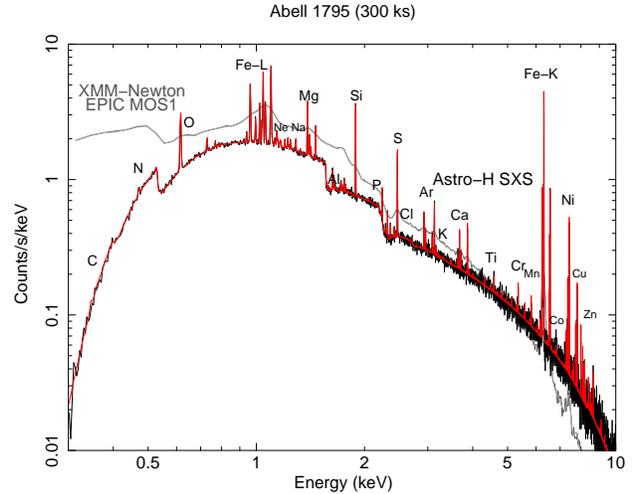}
\end{center}
\vspace{0.2cm}
\caption{Simulated Astro-H micro-calorimeter spectrum of Abell 1795 compared to an XMM-Newton EPIC MOS 
spectrum. The spectra were calculated using an updated version of the atomic database in SPEX 
including trace elements, like Cl, K, Ti, Cr, and Mn.}
\label{fig:sxs}
\end{figure}

\subsection{Future prospects}

The next leap forward will be made using micro-calorimeter technology, which will allow spatially
resolved high-res\-o\-lu\-tion X-ray spectroscopy of extended sources. This type of sensor, which will
be flown on Astro-H (Takahashi et al. 2010\nocite{takahashi2010}), has a typical resolution
of a few eV across the soft X-ray band. With this technology, it will be possible to resolve
lines from less abundant elements, like sodium, chromium, and manganese (see Fig~\ref{fig:sxs}). 
Knowledge about the abundances of these trace elements puts additional constraints on supernova 
models. The Mn/Cr ratio, for example, appears to be very sensitive to the metallicity of the 
type Ia progenitor (Badenes et al. 2008\nocite{badenes2008}) and the Na abundance is, like 
nitrogen, also correlated with the contribution of intermediate mass AGB stars and sensitive to the IMF. 

To measure trace elements like chromium and man\-ga\-nese in high resolution spectra, the spectral 
fitting codes that are used to fit the spectra need updates to their atomic data. Until recently,
only the lines from elements with a high abundance (typically with even atomic numbers) where 
tak\-en into account in the codes. The commonly used databases like ATOMDB\footnote{http://www.atomdb.org} 
(Smith et al. 2001\nocite{smith2001}) and the SPEX\footnote{http://www.sron.nl/spex} line database 
(Kaastra et al. 1996\nocite{kaastra1996}) are being updated and are (partly) 
available in the spectral fitting codes. The systematic errors in the atomic data, which can 
reach the 20\% level in some important lines, remain an issue. Further investments in laboratory
measurements of lines and theoretical line calculations are needed to improve the accuracy.    

Until now, the measured abundances have been compared to individual supernova models. It would, 
however, be much more realistic to estimate the expected a\-bun\-dances from advanced binary
population synthesis codes. Using these codes, it is possible to give the proper weight to
the dif\-fer\-ent type Ia progenitor scenarios and their respective yields. The high calcium 
abundance measured by de Plaa et al. (2007), for example, might be explained by sub-Chandrasekhar
type Ia's, because explosive helium fusion produces more intermediate mass elements, like calcium.
Predicted abundances using these detailed binary population synthesis can be directly compared
to the measured abundances in clusters. The high amount of measurable elements will substantially 
improve the quality of the test and therefore also our knowledge of chemical enrichment in 
clusters.

\acknowledgements The author likes to thank Jelle Kaastra for carefully reading the 
manuscript and useful discussions. SRON Netherlands Institute for Space Research is 
supported financially by NWO, the Netherlands Organisation for Scientific Research.

% Use this code if you wish to generate your bibliography with BibTeX;
% please replace first the string "an-demo" below with the name(s) of
% the BibTeX data base(s) you want to use.
% The resulting bibliography-output (the contents of the .bbl file)
% must be pasted into this file before submission.
% 
\bibliographystyle{an}
\bibliography{jdeplaa}
% 
% Replace the following example bibliography with your references
% before submission:
%\begin{thebibliography}{}
%  \bibitem{} Author1, A.B., Author2, C.D.: 2001, AN 322, 1
%  \bibitem{} Author3, E.F., Author4, G.H.: 2001, AN 322, 10
%  \bibitem{} Author5, I.: 2001, AN 322, 20
%  \bibitem{} Author6, J.: 2001, AN 322, 30
%\end{thebibliography}

\end{document}